# Catheter Ablation Outcome Prediction With Advanced Time-Frequency Features of the Fibrillatory Waves From Patients in Persistent Atrial Fibrillation

Pilar Escribano[1], Juan Ródenas[1], Miguel A Arias[2], Philip Langley[3], José´ J Rieta[4], Raúl Alcaraz[1]

[1] Research Group in Electronic, Biomedical and Telecommunications Engineering,
University of Castilla-La Mancha, Albacete, Spain
[2] Cardiac Arrhythmia Department, Hospital Virgen de la Salud, Toledo, Spain
[3] Faculty of Science and Engineering, University of Hull, Hull, United Kingdomn
[4] BioMIT.org, Electronic Engineering Department, Universitat Politecnica de Valencia, Spain

**Abstract**

*Although catheter ablation (CA) is still the first-line treatment for persistent atrial fibrillation (AF) patients, its limited long-term success rate has motivated clinical interest in preoperative prediction on the procedure's outcome to provide optimized patient selection, limit repeated procedures, hospitalization rates, and treatment costs. To this respect, dominant frequency (DF) and amplitude of fibrillatory waves (f-waves) reflected on the ECG have provided promising results. Hence this work explores the ability of a novel set of frequency and amplitud f-waves features, such as spectral entropy (SE), spectral flatness measure (SFM), and amplitud spectrum area (AMSA), along with DF and normalized f-wave amplitude (NFWA), to improve CA outcome prediction. Despite all single indices reported statistically significant differences between patients who relapsed to AF and those who maintained sinus rhythm after a follow-up of 9 months for 204 6 s-length ECG intervals extracted from 51 persistent AF patients, they obtained a limited discriminant ability ranging between 55 and 62%, which was overcome by 15–23% when NFWA, SE and AMSA were combined. Consequently, this combination of frequency and amplitude features of the f-waves seems to provide new insights about the atrial substrate remodeling, which could be helpful in improving preoperative CA outcome prediction.*

## 1.  Introduction

Atrial fibrillation (AF) is a common supraventricular tachyarrhythmia disrupting sinus rhythm (SR) of the heart. Indeed, it is the most frequently encountered cardiac arrhythmia in clinical practice, roughly affecting 37.5 million people worldwide [1]. Beyond reducing the quality of life, AF often provokes symptoms such as palpitations attacks, dyspnea, inappropriate acceleration of the heart rate, fatigue, chest pain, shortness of breath [1], and, which is more important, AF is the most common risk factor of ischemic stroke [2].

Depending on the duration and recurrent nature of the arrhythmic episodes, AF is classified in several stages that often progresses to sustained forms in a few years [3]. The starting point is usually paroxysmal AF, which is characterized by self-terminates episodes without external intervention within a week. When this is not possible and clinical intervention is required to restore SR, it is classified as persistent. If, despite the attempts to maintain SR, the arrhythmia persists after a year, AF is classified as long-standing persistent. Finally, the most advanced stage of the disease is permanent AF, in which the patient and clinician decide not to make more efforts to stop the arrhythmia due to its strong permanence [3]. Since AF leads to persistent changes in atrial structure and function and then promotes its perpetuation [4], the better way to restore SR should be found as soon as posible [5].

In this context, catheter ablation (CA) is today the first-line strategy for AF treatment [6], which is based on the pulmonary vein isolation (PVI). Despite its short-term effectiveness in most persistent AF patients, the arrhythmia recurs in about 40–50% of them within the first year [6]. This situation explains the emerging clinical interest in preoperative prediction of CA outcome to select those patients who may benefit from the treatment, reducing hospitalization rates, limiting repeated procedures, and reducing treatment costs [7].

So far, some clinical predictors of AF recurrence after CA have been explored, such diabetes, duration of continuous AF, and left atrial size, but they have only provided controversy results [8]. As an alternative, some markers from the electrocardiogram (ECG) recording, such as dominant frequency (DF) computed from the fibrillatory waves ($f$-waves) [9, 10] and the $f$-wave amplitude (FWA) [11, 12], have also been analyzed reporting a promising ability to anticipate CA outcome and being associated with the degree of electrical remodeling presented by the atria. However, as there is still no study analyzing parameters that combine information from both aspects, the aim of this work is to explore a novel $f$-waves set of frequency and amplitude features to improve preoperative outcome prediction of CA.

## 2. Methods

### 2.1. Study population

51 persistent AF patients (9 women and 42 men) constitute the study population who underwent radiofrequency catheter ablation at University Hospital of Toledo, Spain. All antiarrhythmic drug therapy except amiodarone was withheld >5 half lives before the study. The procedure starts with patient sedation using general anesthesia or conscious sedation. Next, catheters were introduced through the femoral venous access, whereas left atrial access was achieved by transeptal puncture. Anticoagulation was maintained through an initial heparin bolus, and further heparin delivery was based on monitoring of activated coagulation time throughout the procedure. PVI consists of the creation of electrically impenetrable boundaries surrounding the ostia of pulmonary veins (PV) and was carried out delivering ablation lesions by placing a catheter and applying radiofrequency current point-by-point for at least 30 seconds to create a contiguous antral circumferential line around each PV, whose location were determined using a mapping catheter. The procedure finished when all PVs were successfully isolated or after restoring SR by electrical cardioversion if AF still remained at the end of the procedure.

The procedure was initially successful in all patients, who did not suffer from any complication after monitoring them for some hours since the intervention finished and during a follow-up of nine months. After this time, 30 patients maintained SR and the remaining 21 relapsed to AF. Note that all patients received anticoagulants and antiarrhythmic drugs by clinical judgment.

### 2.2. Signal acquisition and preprocessing

A standard 12-lead ECG signal was continuously recorded before the ablation with a 977 Hz sampling rate and 16 bits resolution for between 6.4 and 1361.9 seconds. Only lead V1 was analyzed because it reflects the largest $f$-waves compared with the ventricular activity [13]. This lead was resampled to 1 kHz before preprocessing it for removal of baseline wander using a low-pass filtering with cut-off frequency of 0.8 Hz and subtracting the result from the original signal [14]. Moreover, the powerline interference was extracted by means of a wavelet-based denoising algorithm [15], and the remaining high frequency noise was low-pass filtering with cut-off frequency of 70 Hz to obtain a signal as clean as possible[14].

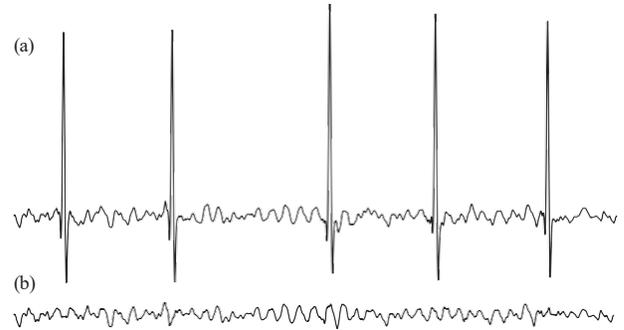

Figure 1. Example of an ECG interval (a) along with the extracted $f$-waves (b).

Afterwards, $f$-waves were extracted from the preprocessed signal by making use of a well-established QRST cancellation method [16]. As an example, Fig. 1 shows the $f$-waves obtained from a typical ECG interval.

### 2.3. Characterization of the $f$-waves

The signal containing the $f$-waves for each patient was divided into non-overlapped 6 s-length excerpts, obtaining 204 segments (84 from patients who relapsed to AF and 120 from those who maintained SR) that were characterized in terms of the parameters described below.

As a reference, DF was obtained for every excerpt as the frequency with the highest power spectral density (PSD) amplitude within the 3–12 Hz range [9, 10]. PSD of the $f$-waves was computed using the Welch Periodogram with a Hamming window of 3.000 points in length, a 50% overlapping between adjacent windowed sections, and a 6.000-points fast Fourier transform (FFT). Also for comparison, the normalized FWA (nFWA) was obtained by expressing the FWA, which is the root mean square value of the $f$-waves [17], as a percentage of the R-peak magnitude to avoid effects that can turn the ECG amplitude higher or lower, such as different gain factors during recording, electrodes impedance, skin conductivity, etc.

In addition to these parameters, three novel indices combining information from time and frequency domains were analyzed, i.e., spectral entropy (SE), spectral flatness measure (SFM), and amplitude spectrum area (AMSA). More precisely, SE quantifies spectral complexity of the $f$-wave signal by computing the sparseness of its spectral distribution [18]. Thus, PSD of the $f$-waves was obtained as for DF and, after its normalization to obtain a probability function with unit area, Shannon entropy was computed between 3 and 12 Hz [18].

On the other hand, SFM provides a measure of how the spectral content of a signal is distributed [19]. It was obtained by dividing geometric and arithmetic means of the $f$-wave spectral distribution. PSD of the $f$-waves was obtained as described before. The index ranges between 0 and 1, so that lower values correspond to higher spectral concentration in small number of frequency bands, and larger values suggest spectral distributions with similar amount of power in all frequency bands [19].

Finally, the index AMSA represents a weighted sum of amplitudes in the spectral domain [20]. It was computed as the sum of the products of individual frequencies and their amplitudes, being PSD of the $f$-waves obtained as in the previous metrics [20].

## 2.4. Performance assessment

For all analyzed indices results were expressed as mean ± standard deviation. Moreover, statistical differences between groups of patients were tested by a Student's $t$-test. A two-tailed value of $p < 0.05$ was considered as statistically significant.

On the other hand, the ability of each feature to discriminate between both group of patients was evaluated by means of a receiver operating characteristic (ROC) curve. This plot is the result of plotting the fraction of true positives out of positives (sensitivity), which is considered as the percentage of patients who relapsed to AF correctly classified, against the fraction of false positives out of the negatives (1−specificity), which is the rate of patients maintaining SR, at various threshold settings. The optimal threshold was selected as the one providing the best balance between Se and Sp, although in this way the highest percentage of patients correctly classified, i.e., accuracy (Acc), could not be achieved. Finally, the area under the ROC curve (AROC) was also obtained as an aggregate measure of performance of a variable across all possible classification thresholds.

To improve classification, a linear discriminant analysis (LDA) was also performed. Variable selection was carried out by a forward stepwise approach including, at each step, the feature which led to maximization of the Lawley–Hotelling trace (Rao's V).

## 3. Results

The results of the single indices computed from the 204 6 s-length ECG excerpts are shown in Table 1. As can be observed, all of them reported statistically significant differences between both groups of patients, obtaining $p$-values lower than 0.05. Moreover, whereas DF and AMSA provided higher mean values for the patients who relapsed to AF, the remaining indices reported lower values than for those maintaining SR during the follow-up.

Table 1. Mean and standard deviation values for the analyzed metrics from both group of patients.

| Index | Group of patients | | $p$-value |
|---|---|---|---|
| | maintaining SR | relapsing to AF | |
| DF (Hz) | 6.01 ± 1.47 | 6.44 ± 1.25 | 0.05 |
| nFWA (%) | 6.47 ± 4.45 | 5.14 ± 2.98 | 0.04 |
| SE (no units) | 0.77 ± 0.08 | 0.74 ± 0.08 | < 0.01 |
| SFM (%) | 37.1 ± 13.8 | 32.6 ± 13.8 | 0.012 |
| AMSA ($\mu$V · Hz) | 189.3 ± 82.7 | 231.8 ± 85.7 | < 0.01 |

Table 2. Classification results obtained by the parameters computed from the $f$-waves.

| Index | Se (%) | Sp (%) | Acc (%) | AROC (%) |
|---|---|---|---|---|
| DF | 57.14 | 57.27 | 57.21 | 57.38 |
| nFWA | 55.45 | 54.76 | 55.15 | 58.63 |
| SE | 61.82 | 61.9 | 61.85 | 63.76 |
| SFM | 60.00 | 60.71 | 60.31 | 60.55 |
| AMSA | 60.71 | 60.91 | 60.82 | 64.56 |

Regarding classification between patients, Table 2 shows values of Se, Sp, Acc and AROC achieved by all single metrics. Only limited values ranging between 55 and 65% were noticed. Nonetheless, the combination of nFWA, SE and AMSA through a LDA reached improvements of 15–23% with respect to the single indices, since values of Se, Sp, Acc and AROC of 77.38, 77.27, 77.32, and 79.79% were obtained, respectively.

## 4. Discussion and conclusions

Clinical interest in CA outcome prediction has emerged due to its high rate of long-term inefficiency in persistent AF patients. The main goal is optimizing patient selection and then enabling tailored approaches, reducing treatment costs, and limiting associated risk [7]. For that purpose, advanced time-frequency characterization of the $f$-waves in terms of SE, SFM and AMSA has been for the first time addressed in the present work. The three indices have shown a better correlation with the outcome of CA than other common metrics, such as DF and nFWA. In fact, they have exhibited improvements between 2 and 8% in values of Acc and AROC. Moreover, the LDA-based combination of some of these metrics also increased diagnostic accuracy by more than 15% regarding all single parameters.

The mean values obtained by these indices also agree with previous findings. Thus, SE and SFM presented higher values for the patients maintaining SR than for those relapsing to AF during the follow-up, thus suggesting a more uniform spectral distribution of the $f$-waves. In line with this result, previous works have revealed that the presence of organized $f$-waves is indicative of a higher likelihood of spontaneous termination of paroxysmal AF, as well as of a successful outcome in electrical cardioversion [17]. On the other hand, AMSA presented higher values for the patients who relapsed to AF than for those maintaining SR. Because this index is computed as the sum of the product of individual frequencies and their amplitudes, frequency seems to predominate over amplitude. This outcome is consistent with the finding that DF and its first harmonic were more predictive of the mid-term CA success than their amplitudes [9]. Thus, this work has provided clinically useful information about the long-term outcome of CA in persistent AF, providing new insights about the atrial substrate remodeling before the procedure. Nonetheless, further studies with wider databases will have to be conducted in the future to corroborate these results.

## Acknowledgment

This research has been supported by grants DPI2017-83952-C3 from MINECO/AEI/FEDER EU, SBPLY/17/180501/000411 from Junta de Comunidades de Castilla-la Mancha and AICO/2019/036 from Generalitat Valenciana. Moreover, Pilar Escribano holds a graduate research scholarship from University of Castilla-La Mancha.

Address for correspondence:

Pilar Escribano Cano
E.S.I. Informática., Campus Univ., 02071, Albacete, Spain
e-mail: pilar.escribano1@alu.uclm.es